\documentclass{edp-conf}
\usepackage{graphicx}
\usepackage{rotate}
\usepackage{amsmath}
\usepackage{amsfonts}
\usepackage{graphicx}
\usepackage{graphics}
\usepackage{epsfig}
\begin{document}

\TitreGlobal{SF2A 2005}

\title{DETECTING CHAOTIC AND ORDERED MOTION IN BARRED GALAXIES}
\author{MANOS, T.}\address{Observatoire Astronomique de Marseille-Provence
(OAMP), FRANCE and Center for Research and Applications of Nonlinear
Systems (CRANS), Department of Mathematics, University of Patras,
GREECE.}
\author{ATHANASSOULA, E.}\address{Observatoire Astronomique de Marseille-Provence
(OAMP), FRANCE.}
%
%
\setcounter{page}{237}
\index{MANOS, T.} \index{ATHANASSOULA, E.}

\maketitle
\begin{abstract}
A very important issue in the area of galactic dynamics is the
detection of chaotic and ordered motion inside galaxies. In order to
achieve this target, we use the Smaller ALignment Index (SALI)
method, which is a very suitable tool for this kind of problems.
Here, we apply this index to 3D barred galaxy potentials and we
present some results on the chaotic behavior of the model when its
main parameters vary.
 \end{abstract}
%
\section{Introduction}
The Smaller ALignment Index (SALI) (Skokos 2001; Skokos et al. 2004)
or, as elsewhere called, Alignment Index (AI) (Voglis et al. 2002),
can distinguish between chaotic and ordered motion in dynamical
systems. In order to compute the SALI for a given orbit, one has to
follow the time evolution of the orbit itself and of two deviation
vectors $v_{1}$ and $v_{2}$, which initially point in two different
directions. At every time step the two deviation vectors are
normalized and the SALI is then computed as:
\begin{equation}\label{eq:1}
    SALI(t)=min\{\|\frac{v_{1}(t)}{\|v_{1}(t)\|}+\frac{v_{2}(t)}{\|v_{2}(t)\|}\|,\|\frac{v_{1}(t)}{\|v_{1}(t)\|}-\frac{v_{2}(t)}{\|v_{2}(t)\|}\|\}.
\end{equation}
The SALI for a chaotic trajectory tends to zero, while for a regular
one it fluctuates around a positive number.

\section{Discussion of the applications and the results in 3 dof Ferrers}
We apply the method to a 3D Ferrers potential \cite{} which consists
of the superposition of a Miyamoto sphere, a Plummer disc and a
Ferrers bar.
\begin{figure}[h]
   \centering
\includegraphics[width=3.8cm]{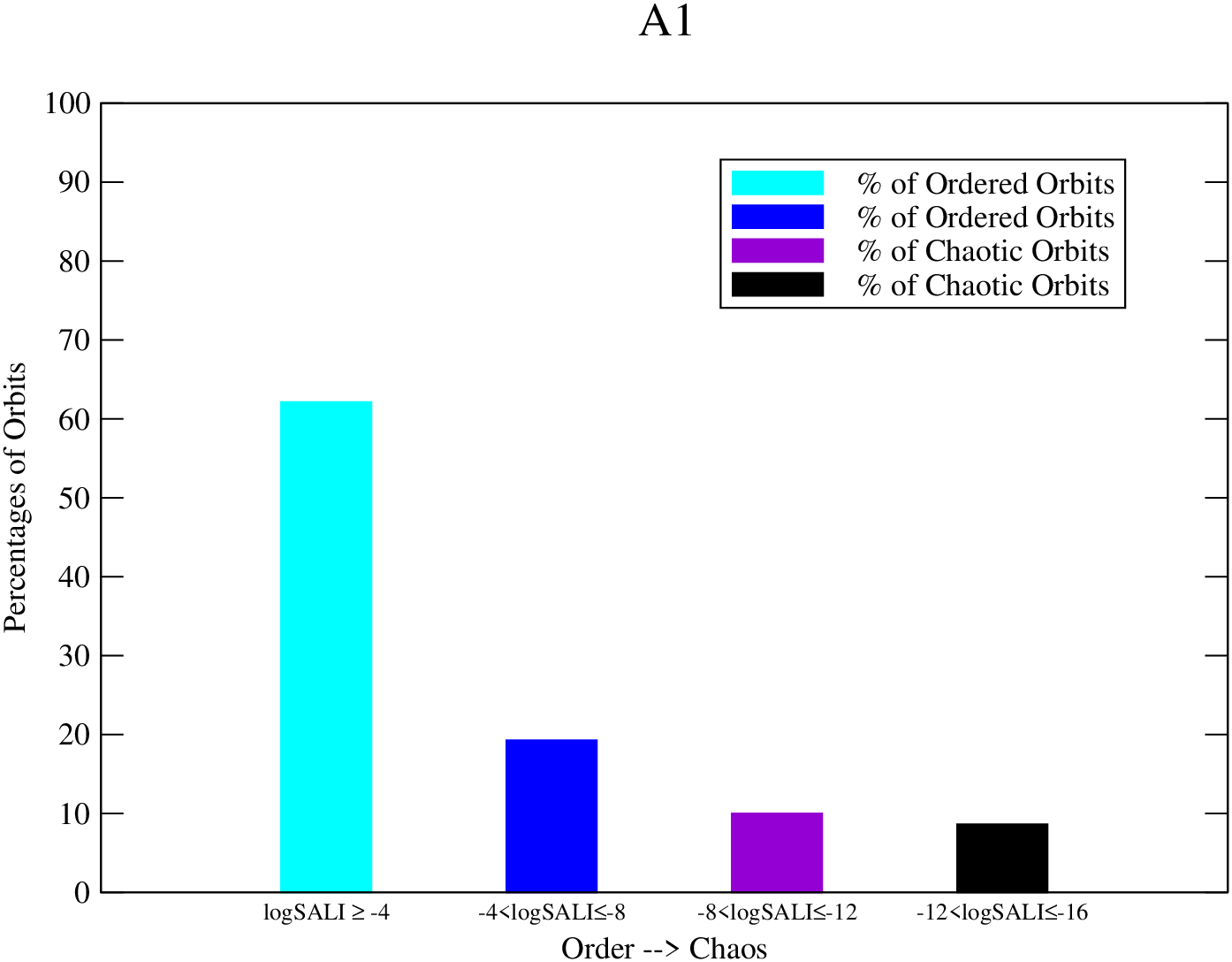}\hspace{0.1cm}
\includegraphics[width=3.8cm]{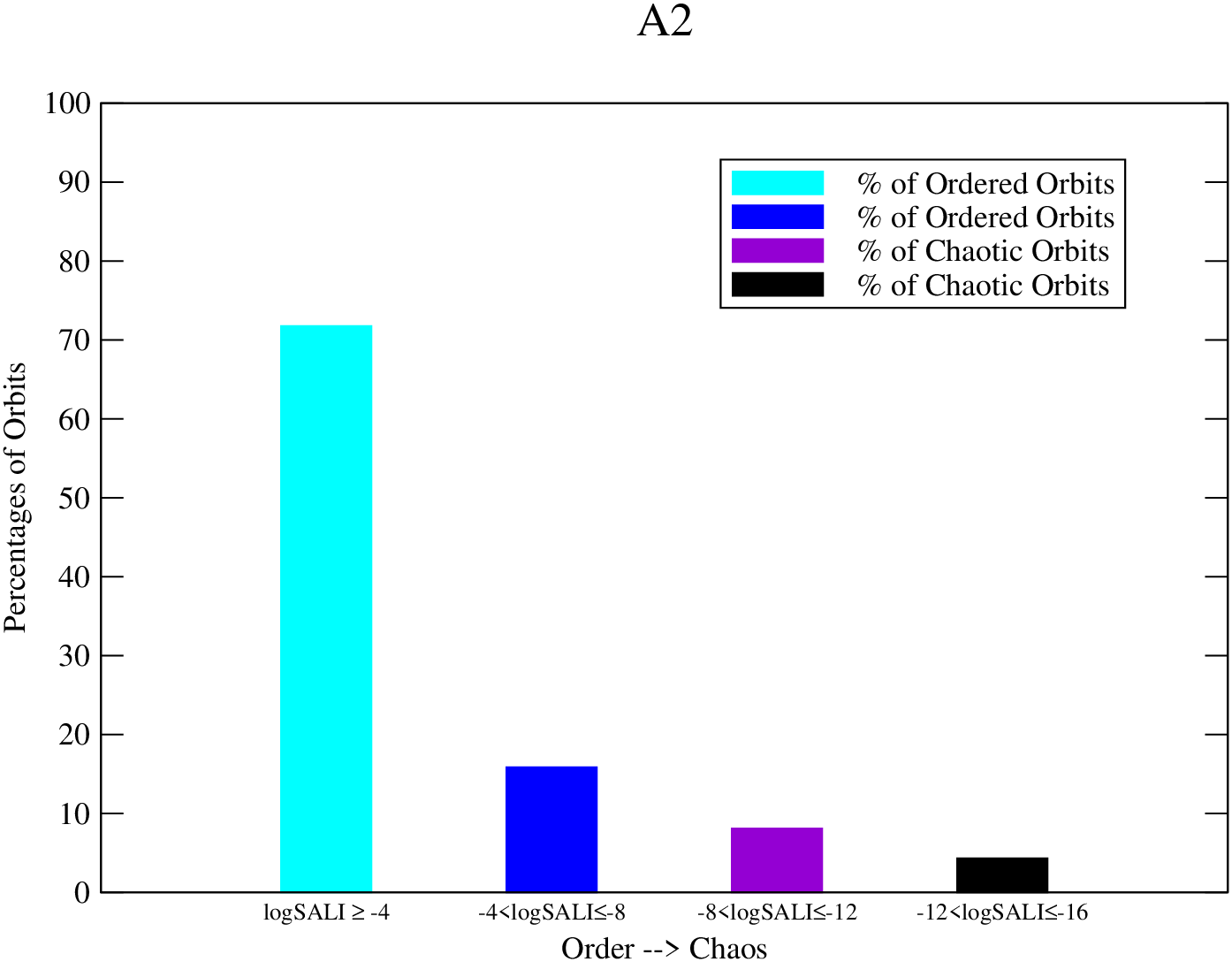}\hspace{0.1cm}
\includegraphics[width=3.8cm]{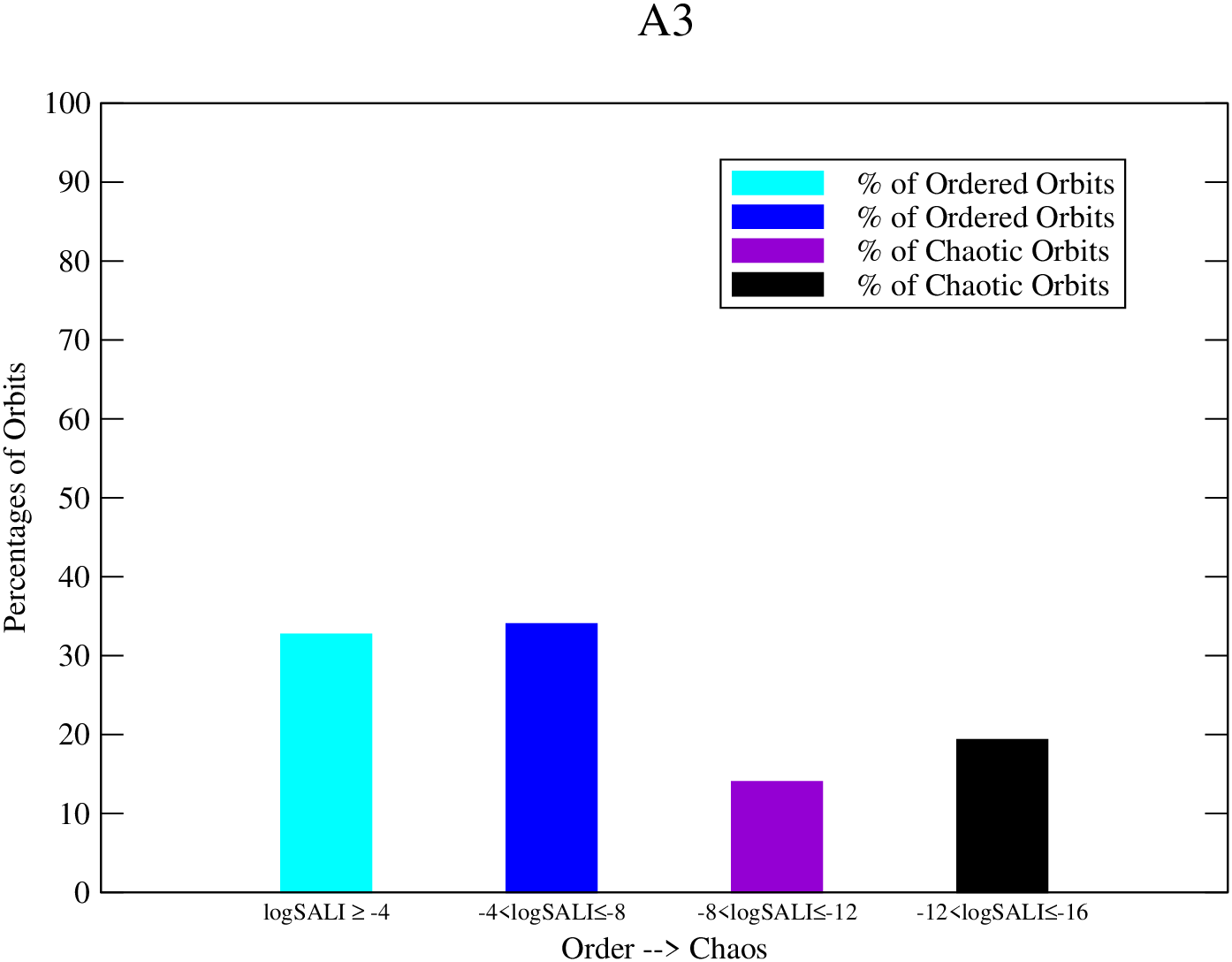}\\\vspace{0.3cm}
\includegraphics[width=3.8cm]{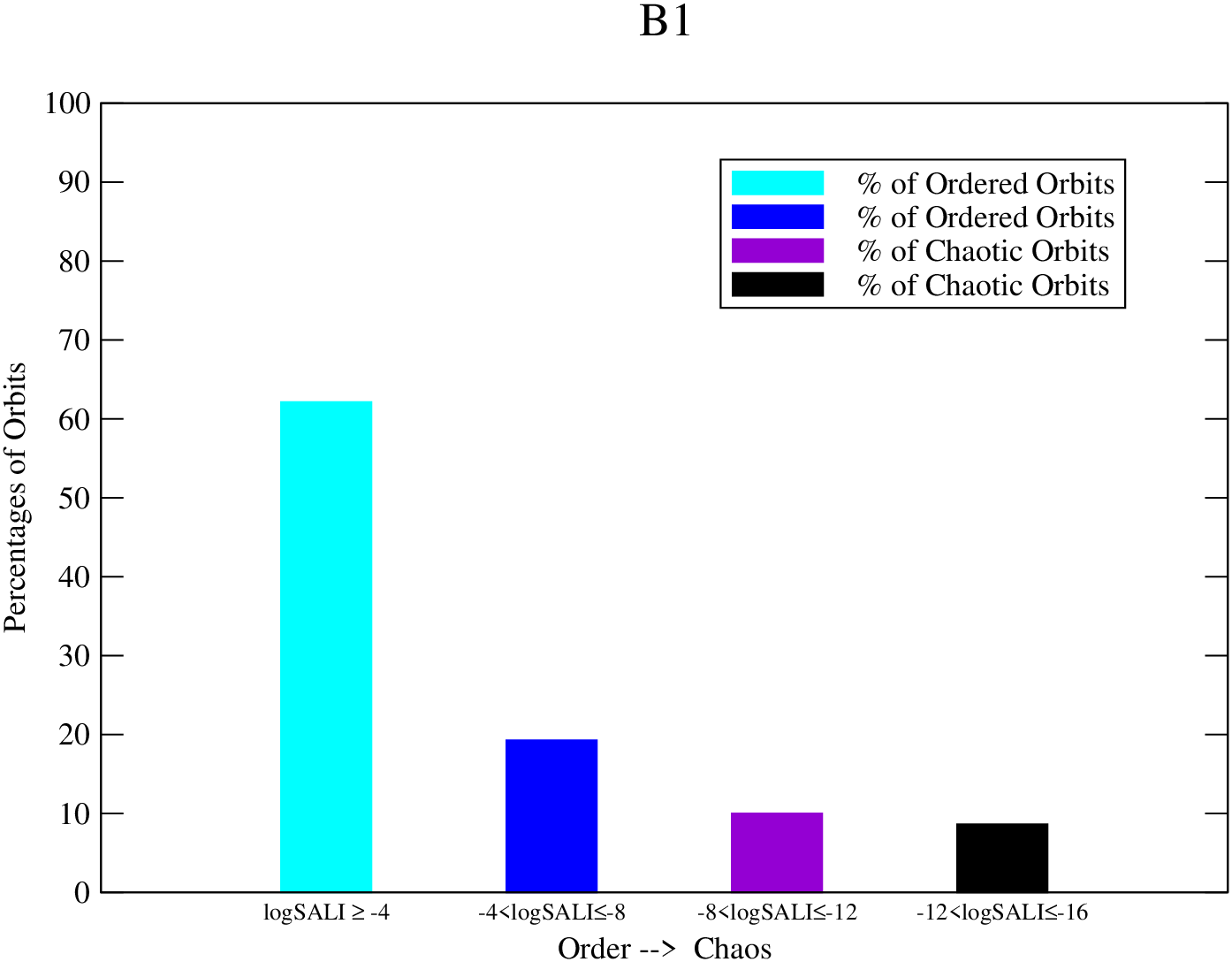}\hspace{0.1cm}
\includegraphics[width=3.8cm]{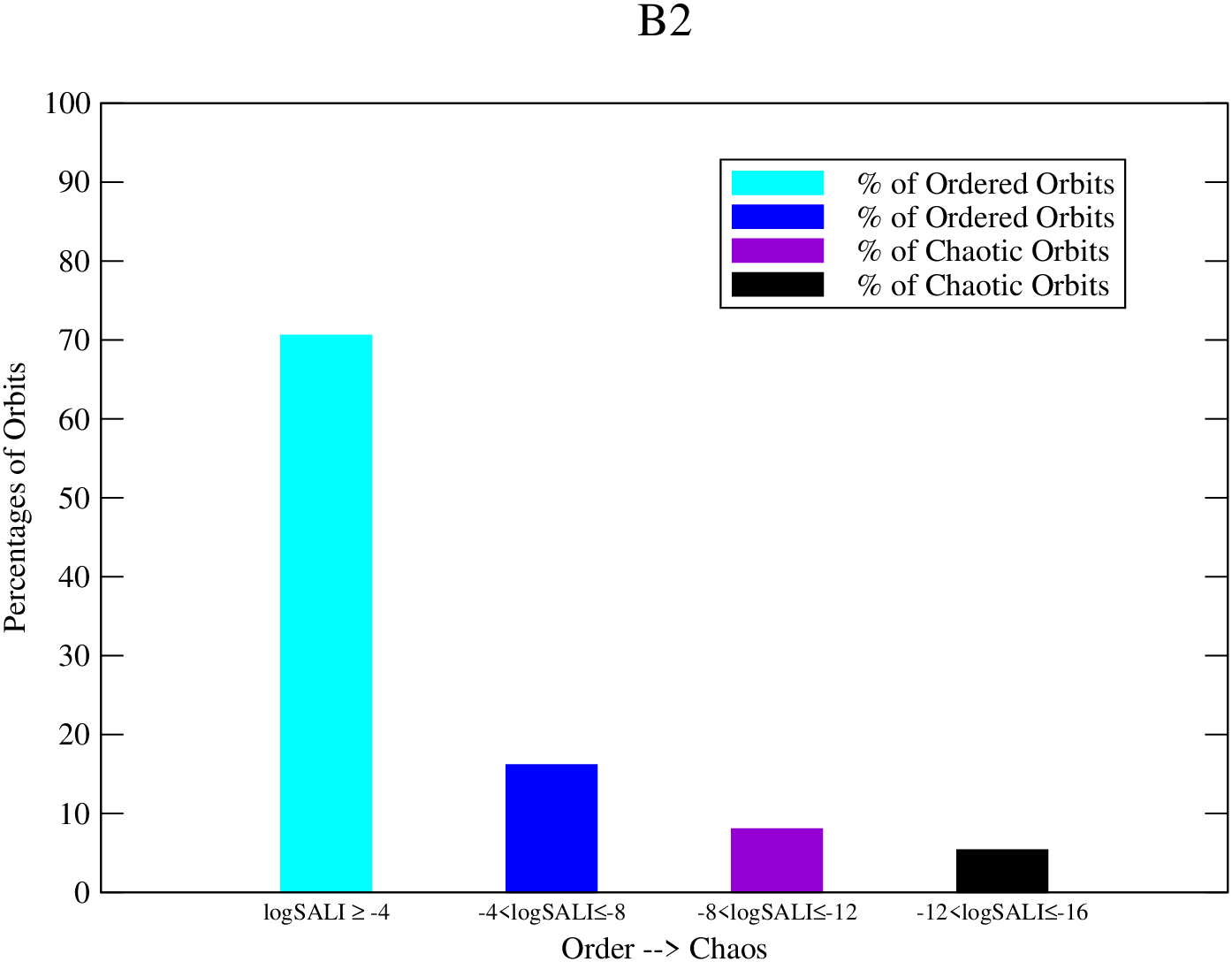}\hspace{0.1cm}
\includegraphics[width=3.8cm]{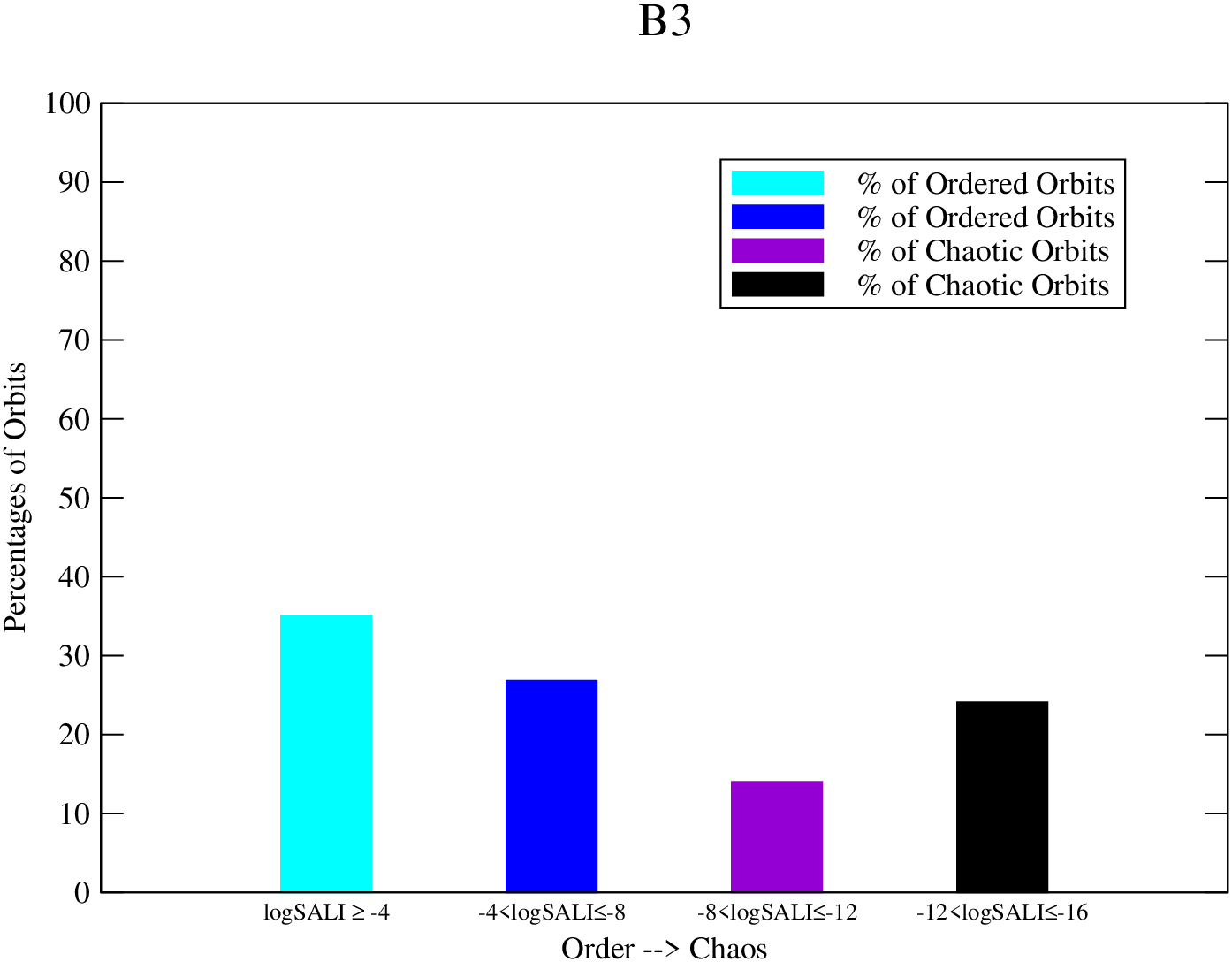}
      \caption{Percentages of regular (first and second bar) and chaotic (third and forth bar) orbits, for our standard model
      (left column of panels), a model with a thick bar (central column) and a model with a massive bar (right column).
      The two rows show two different ways of choosing the orbital population.}
       \label{figure_mafig}
   \end{figure}
In figure 1 we present percentages of chaotic, intermediate and
regular trajectories, where we vary the mass of the bar component
(panels A3-B3) and the length of the short $z$-axis (panels A2-B2)
of the initial models A1 and B1. The two rows differ in the way we
give the 27000 initial conditions. For the A1, we give initial
conditions in the plane $(x,p_{y},z)$ with $(y,p_{x},p_{z})=(0,0,0)$
and for the B1, in the plane $(x,p_{y},p_{z})$ with
$(y,z,p_{z})=(0,0,0)$. By comparing the results, we see that the
increase of the bar mass causes more chaotic behavior in both cases
(panels A3, B3). This confirms the results by Athanassoula et al.
(1983) in 2 dof. On the other hand, it is obvious that when the bar
is thicker, i.e. the length of the $z$-axis larger, the system gets
more regular. We also, investigated how the pattern speed of the bar
affects the system and found that the percentage of the regular
orbits is greater in slow bars.\\









\textbf{Acknowledgements.} Thanos Manos was partially supported by
"Karatheodory" graduate student fellowship No  B395 of the
University of Patras and by "Marie –- Curie" fellowship No
HPMT-CT-2001-00338.

\end{document}